\documentclass{appolb}
\usepackage{epsfig}
\usepackage{axodraw}
\usepackage{amssymb}
\usepackage{amsmath}
\usepackage{cite}

\newcommand{\ds}{\displaystyle}
\newcommand{\eps}{\varepsilon}

\begin{document}

\title{
\vspace*{-3mm}
{\small DESY 07-096 \hfill SFB/CPP-07-71}\\
Heavy Flavour Production in Deep--Inelastic Scattering -
    Two--Loop Massive Operator Matrix Elements and Beyond%
\thanks{Presented at the XXXI International Conference of Theoretical Physics: Matter To The Deepest, Ustron, Poland, 5-11 September 2007}
}
\author{
I.~Bierenbaum, J.~Bl\"umlein and S.~Klein
\address{Deutsches Elektronen-Synchrotron, DESY,
Platanenallee 6, D-15738 Zeuthen, Germany}
}
\maketitle
\begin{abstract}
\noindent
  We calculate the O($\eps$)--term of the two--loop massive operator
  matrix elements for twist 2--operators, which contribute to the heavy
  flavour Wilson coefficients in unpolarized deep--inelastic
  scattering in the asymptotic limit $Q^2 \gg m^2.$ Our calculation
  was performed in Mellin space using Mellin--Barnes integrals and
  generalized hypergeometric functions. The O($\eps$)--term contributes
  in the renormalization at 3--loop order.
\end{abstract}

\PACS{PACS numbers come here}
  
\section{Introduction}
\noindent
In the range of small values of the Bj$\o$rken variable $x$, the
contribution of heavy flavour corrections to deep--inelastic structure
functions is of the order of 20--40\% and hence has to be known in the
QCD analyses of the structure functions for high
precision extractions of the parton densities and the QCD scale 
$\Lambda_{\rm QCD}$~\cite{LAM}.
In the full kinematic 
range, a semi--analytic result for
the heavy flavour Wilson coefficients up to next--to--leading order
exists \cite{Laenen:1992zk}, with a fast implementation to
Mellin--space given in \cite{Alekhin:2003ev}, whereas a fully analytic
result to O($\alpha_s^2$) could be achieved in the limit $Q^2 \gg m^2$
in \cite{Buza:1995ie}, $Q^2$ denoting the virtuality of the exchanged
photon and $m^2$ the mass of the heavy quark. The corresponding Wilson 
coefficient for $F_L$ at $O(\alpha_s^3)$ was calculated in \cite{FL}. In 
this limit, the
heavy--flavour contributions can be expressed as a convolution of
light--flavour Wilson coefficients and massive operator matrix
elements (OMEs) between light partonic states. The results in
\cite{Buza:1995ie} have been obtained using integration--by--parts
techniques. We performed a first recalculation of these OMEs in
Mellin--space \cite{Bierenbaum:2007dm,Bierenbaum:2007qe}, using both
Mellin--Barnes integrals and generalized hypergeometric functions.
This shifts the problem of solving complicated integrals of
Nielsen--type in \cite{Buza:1995ie}, to the calculation of sums over
products of harmonic sums \cite{Vermaseren:1998uu,Blumlein:1998if}
depending on the Mellin--parameter $N$, weighted binomials and
Euler Beta--functions. The expressions in our result are even on
the diagrammatic level considerably smaller than the ones obtained in
\cite{Buza:1995ie}, and are seemingly more suitable to the problem.
In this paper, we show a first step towards the O($\alpha_s^3$)--term
of the heavy--flavour Wilson coefficients, by calculating in
dimensional regularization the O($\eps$)--term of the two--loop OMEs,
with $\eps=D-4$.

\section{Method}
\noindent
Our calculation is performed in the asymptotic limit $Q^2 \gg m^2$,
applying the light--cone expansion, where, as the massless renormalization
group equation (RGE) gives a splitting of the deep--inelastic
structure functions $F_{2/L}$ into a convolution of perturbatively
calculable Wilson coefficients and non--perturbative parton
distribution functions, the massive RGE allows to write the heavy
flavour contribution to the twist--2 Wilson coefficients as a
convolution of light--flavour Wilson coefficients and massive operator
matrix elements \cite{Buza:1995ie}:
    \begin{eqnarray}
      {H_{(2,L),i}^{{\rm S,NS}}\left(\frac{Q^2}{\mu^2},
            \frac{m^2}{\mu^2}\right)} 
       = \underbrace{{A_{k,i}^{{\rm S,NS}} \left(\frac{m^2}{\mu^2}\right)}}_
	{
	\begin{array}{l}
	\mbox{massive OMEs}
	\end{array}
	} 
	\otimes \underbrace{{C_{(2,L),k}^{{\rm S,NS}} 
                \left(\frac{Q^2}{\mu^2}\right)}~.}_{
	\begin{array}{l}
          \mbox{light Wilson coefficients}
	\end{array}
        } \nonumber
    \end{eqnarray}
    These OMEs are universal objects, calculable via the corresponding
    flavour singlet, pure--singlet and non--singlet operators between
    partonic states, determining the non--power contributions in
    $m^2/Q^2$. The process dependence is then solely given by the
    massless light Wilson coefficients \cite{WILS}.\\ The OMEs contain
    ultraviolet and collinear divergences, the former being removed through
    renormalization, the latter absorbed into the parton distribution
    functions. To two--loop order, the renormalized gluonic OME
    reads:
\begin{eqnarray}
     {A_{Qg}^{(2)}} &=& \frac{1}{8}\left\{ {\widehat{P}_{qg}^{(0)}} 
                            \otimes 
                            \left[{P_{qq}^{(0)}} - {P_{gg}^{(0)}} 
                          + 2 \beta_0\right]\right\} {\ln^2\left(
                            \frac{m^2}{\mu^2}\right)}
                          - \frac{1}{2}{\widehat{P}_{qg}^{(1)}} 
                            {\ln \left(\frac{m^2}{\mu^2}\right)}\nonumber\\
                         & & +{\overline{a}_{Qg}^{(1)}} \otimes
                              \left[{P_{qq}^{(0)}}
                             -{P_{gg}^{(0)}} + 2 \beta_0\right]  
                             + {a_{Qg}^{(2)}}~, \nonumber
\end{eqnarray}
with similar expressions for the quarkonic contributions. Here,
$P_{ij}^{(k-1)}$ are the $k$th--loop splitting functions, $\beta_0$ is
the lowest order expansion coefficient of the $\beta$--function, and
$\mu^2$ the renormalization and factorization scale. ${a}_{ij}^{(k)}$
and $\bar{a}_{ij}^{(k)}$ are the $O(\varepsilon^0)$ resp.
$O(\varepsilon)$-terms in the expansion of the OME. As a first step
towards a O($\alpha_s^3$) calculation, one needs each of these
quantities to one additional order in $\eps$, since they then enter the
constant term of the OME by multiplying the corresponding splitting
functions. The O($\eps$)--term of the OME  $A_{ij}^{(2)}$,
$\bar{a}_{ij}^{(2)}$, is a new result presented here and the main
topic of this calculation.

\vspace*{-5mm}
\section{Calculation}
\noindent\normalsize
The diagrams can be grouped into two sets: one--loop in one--loop
insertions and generic two--loop diagrams. They are calculated using 
{\tt FORM} \cite{FORM} and {\tt MAPLE} programs. Figure 
\ref{fig:examples} shows some diagrams contributing to the gluonic OME.

{
\begin{figure}[t]
\begin{center}
\begin{tabular}{lr}
  \raisebox{0.5cm}{
  \large{b}:
  }
  \quad  \includegraphics[width=2.8cm]{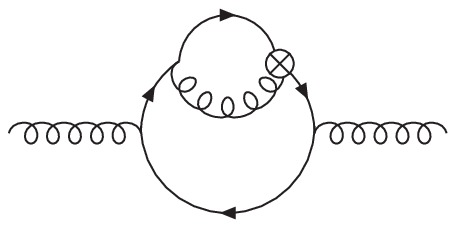} 
&
  \qquad 
  \raisebox{0.5cm}{
  \large{e}:
  }
  \quad \includegraphics[width=2.8cm]{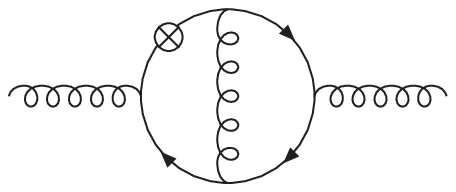}
\end{tabular}
\end{center}
\caption{\small Two example diagrams contributing to the gluonic OME,
  with all fermion lines massive and external momentum $p^2=0$.}
\label{fig:examples}
\end{figure}
}

{
\begin{table*}[b]
\caption{Mellin moments N$=2$ and N$=6$ for diagram $e$.
}
\begin{center}
\includegraphics[width=2cm,angle=90]{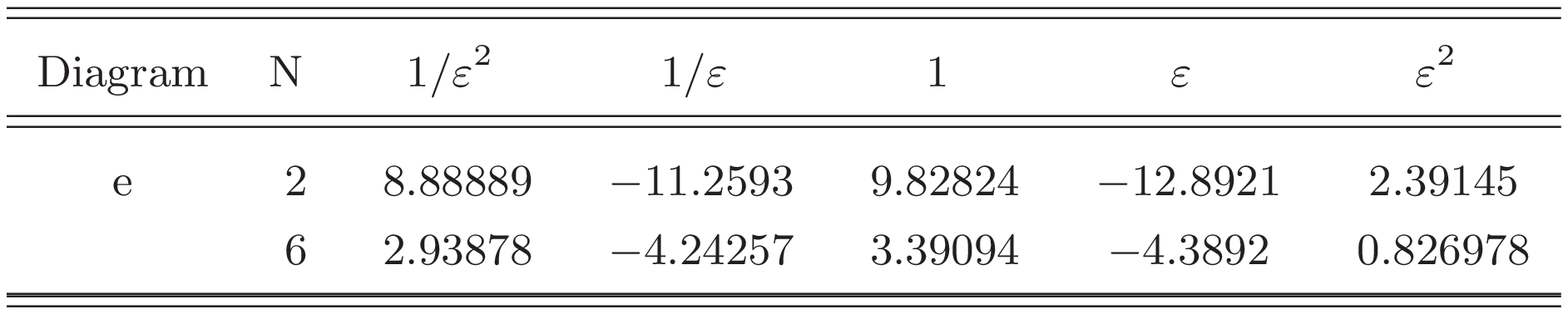}
\end{center}
\label{table:results}
\end{table*}
}

\normalsize
The rules for operator insertion are, e.g., given in
\cite{Floratos:1977au}. The calculation is done on the one hand by
the use of Mellin--Barnes integrals to produce numeric results. These
results serve as a check for the analytic results, obtained by
expressing the diagrams as generalized hypergeometric functions which
are first expanded in $\eps$ and then summed up to the desired
order.\\ The application of Mellin--Barnes integrals for scalar
diagrams in our framework has already been explained in some detail
before, cf. e.g. \cite{Bierenbaum:2007dm} (see also
\cite{Bierenbaum:2003ud}).  The idea is to express in a loop--by--loop
manner the sub--diagram of a full diagram into a Mellin--Barnes
representation and to combine this with the remaining part. For full
diagrams with a numerator structure, one can make heavily use of the
fact that the light--like vector $\Delta$ occurring in the numerators
obeys $\Delta^2=0$. This reduces the integrals to be
calculated to a smaller set. After finding a suitable
Mellin--Barnes integral representation, we use the mathematica package
{\tt MB}~\cite{Czakon:2005rk} to numerically calculate the
Mellin--Barnes integrals for fixed values of Mellin--$N$, up to a
given order in $\eps$. As an example, Table~\ref{table:results} shows
up to some generic multiplicative factors the results for the full
diagram $e$ of Figure \ref{fig:examples}.

\normalsize
By closing the contour and applying the theorem of residues, it is
in principle possible to even obtain analytic results from the
Mellin--Barnes representation \cite{Bierenbaum:2007dm}.  However, the
way of hypergeometric functions turned out to be more appropriate for
the calculation of analytic results. In this case, one first
introduces Feynman parameters and does the two momentum integrations.
As an example, the scalar version of diagram $e$ of Figure
\ref{fig:examples} with all propagators to the power one, evaluates to:
{
  \begin{flalign*}
   I_{e}:&=\frac{(\Delta p)^{{N-1}}\Gamma(1-\eps)}{N(N+1)(4\pi)^{4+\eps}
        (m^2)^{1-\eps}}\\
   &\int_0^1dz\,dw\, \frac{w^{-1-\eps/2}(1-z)^{\eps/2}z^{-\eps/2}}
        {(z+w-wz)^{1-\eps}}\Biggl[1-w^{{N+1}}-(1-w)^{{N+1}}\Biggr]~.
  \end{flalign*}}
\normalsize
  
\noindent
On rewriting this Feynman parameter integral into a
  $\empty_{P}F_Q$--function, one obtains a product of
  $\empty_{3}F_2$--functions and the Euler Beta--function\linebreak
  $B(a,b):=\Gamma(a)\Gamma(b)/\Gamma(a+b)$:
{  
\begin{flalign*}
   I_{e}&=\frac{c}{N(N+1)}
        \exp\Biggl\{\sum_{i=2}^{\infty}\zeta_i\frac{\eps^i}{i}\Biggr\}
        \Biggl\{
        \\&
        B(\eps/2+1,1-\eps/2)B(1,-\eps/2)\:
        {\empty_{3}F_2
        \left[
          \begin{array}{l}
            1-\eps,1,1+\eps/2\\2,1-\eps/2
          \end{array}
          ;1
        \right]}\\
        &
        -B(\eps/2+1,1-\eps/2)B(1,N+1-\eps/2)\:
        {\empty_{3}F_2
        \left[
          \begin{array}{l}
            1-\eps,1,1+\eps/2\\
            2,N+2-\eps/2
          \end{array}
          ;1
        \right]}\\
        &-B(\eps/2+1,1-\eps/2)B(N+2,-\eps/2)\:
        {\empty_{3}F_2
        \left[
          \begin{array}{l}
            1-\eps,N+2,1+\eps/2\\
            2,N+2-\eps/2
          \end{array}
          ;1
        \right]}
         \Biggr\}\\
	&c:=\frac{S^2_{\eps}}{(4\pi)^{4}(m^2)^{1-\eps}}
    (\Delta p)^{N-1}
  \end{flalign*}
}
  
\noindent
One then expands this expression up to the desired order in $\eps$,
  in this obtaining finite and infinite sums over harmonic sums and
  Beta--functions:
  \begin{flalign*}
   I_{e}
   =
    &\frac{c}{N(N+1)} \sum_{s=1}^{\infty}\Biggl\{
        \frac{1}{s^2}-\frac{S_1(s)}{s}+\frac{S_1(N+s)}{s}
        -\frac{B(N+1,s)}{s}
         \Biggr\}
         +O(\eps)
  \end{flalign*}
  It is the next step of summing up the various sums over harmonic sums
  and more complicated expressions, which constitutes the most
  difficult part of the calculation. These sums could be solved among
  other things using their integral representations, where a certain
  amount of more complicated sums could be calculated using the mathematica package {\tt SIGMA}
  \cite{SCHNEIDER,Bierenbaum:2007zu}. For integral $e$, one obtains up
  to O($\eps$):
  \begin{flalign*}
   I_{e}=&
	 \frac{c}{N(N+1)}
        \Biggl\{
        \frac{{{S}_1^2({N})}+3{{S}_2({N})}}{2}
        +
        \frac{{{S}_1^3({N})}
          +3{{S}_1({N}){S}_2({N})}
          +8{{S}_3({N})}}{12}\: {\eps}
        \Biggr\}
  \end{flalign*}
  In a similar manner, it was possible to calculate all diagrams
  contributing to the calculation of the O($\eps$)--term of the
  two--loop unpolarized OMEs. Here algebraic relations between harmonic 
sums were used \cite{ALGEBRA}.

\section{Results}
\noindent
The O($\eps$) contributions to the mass--renormalized unpolarized OMEs for 
the
singlet, pure--singlet, and
non--singlet cases read:\\[1em]
%
%
{\footnotesize
\begin{eqnarray}
 {\overline{a}_{Qg}^{(2)}}&=&
           T_R C_F\Biggl\{
              \frac{2}{3}\frac{\ds (N^2+N+2)(3N^2+3N+2)}
                                 {\ds N^2(N+1)^2(N+2)}{\zeta_3}
                +\frac{P_{1}}
                                 {N^3(N+1)^3(N+2)}{S_2} \nonumber\\
 &&
              +\frac{N^4-5N^3-32N^2-18N-4}
                                 {N^2(N+1)^2(N+2)}{S^2_1}
                      +\frac{N^2+N+2}
                           {N(N+1)(N+2)}
                   \Bigl(
                     16{S_{2,1,1}}
                    -8{S_{3,1}}
\nonumber\\ &&
                    -8{S_{2,1}S_1}
                    +3{S_4}  
                    -\frac{4}{3}{S_3S_1}
                    -\frac{1}{2}{S^2_2}
                    -{S_2S^2_1}
                    -\frac{1}{6}{S^4_1}
                    +2{\zeta_2}{S_2}
                    -2{\zeta_2}{S^2_1}
                    -\frac{8}{3}{\zeta_3}{S_1}
                   \Bigr) \nonumber\\ & & 
                -8\frac{N^2-3N-2}
                       {N^2(N+1)(N+2)}{S_{2,1}}
                +\frac{2}{3}\frac{3N+2}
                         {N^2(N+2)}{S^3_1}
\nonumber\\ & &
              +\frac{2}{3}\frac{3N^4+48N^3+43N^2-22N-8}
                         {N^2(N+1)^2(N+2)}{S_3}
                +2\frac{3N+2}
                       {N^2(N+2)}{S_2S_1}
                +4\frac{{S_1}}
                       {N^2}{\zeta_2}
\nonumber\\ & &
                +\frac{N^5+N^4-8N^3-5N^2-3N-2}
                                 {N^3(N+1)^3}{\zeta_2}
\nonumber\\ & &              
              -2\frac{2N^5-2N^4-11N^3-19N^2-44N-12}
                                 {N^2(N+1)^3(N+2)}{S_1}
  +\frac{P_{2}}
                      {N^5(N+1)^5(N+2)}
                          \Biggr\}\nonumber\\
& &      +    T_RC_A\Biggl\{ 
                   \frac{N^2+N+2}
                         {N(N+1)(N+2)}
                      \Bigl(
                        16{S_{-2,1,1}}
                       -4{S_{2,1,1}}
                       -8{S_{-3,1}}
                       -8{S_{-2,2}}
                       -4{S_{3,1}}
                       -\frac{2}{3}{\beta'''}
\nonumber\\
& &
                       +9{S_4}
                     -16{S_{-2,1}S_1}
                       +\frac{40}{3}{S_1S_3}
                       +4{\beta''}{S_1}
                       -8{\beta'}{S_2}
                       +\frac{1}{2}{S^2_2}
                       -8{\beta'}{S^2_1}
                       +5{S^2_1S_2}
                       +\frac{1}{6}{S^4_1}
\nonumber\\ & &
                       -\frac{10}{3}{S_1}{\zeta_3}
                       -2{S_2}{\zeta_2}
                     -2{S^2_1}{\zeta_2}
                       -4{\beta'}{\zeta_2}
                       -\frac{17}{5}{\zeta_2^2}
                     \Bigr)\nonumber\\
&&                  +\frac{4(N^2-N-4)}
                         {(N+1)^2(N+2)^2}
                      \Bigl(
                       -4{S_{-2,1}}
                       +{\beta''}
                       -4{\beta'}{S_1}
                     \Bigr)
                 -\frac{2}{3}\frac{N^3+8N^2+11N+2}
                            {N(N+1)^2(N+2)^2}{S^3_1}
\nonumber\\ & &
                   +8\frac{N^4+2N^3+7N^2+22N+20}
                          {(N+1)^3(N+2)^3}{\beta'}
+2\frac{3N^3-12N^2-27N-2}
                          {N(N+1)^2(N+2)^2}{S_2S_1}
\nonumber
\end{eqnarray}}
{\footnotesize\begin{eqnarray}
&&
                 -\frac{16}{3}\frac{N^5+10N^4+9N^3+3N^2+7N+6}
                             {(N-1)N^2(N+1)^2(N+2)^2}{S_3}
                   -8\frac{N^2+N-1}
                          {(N+1)^2(N+2)^2}{\zeta_2}{S_1}\nonumber\\
&&
                 -\frac{2}{3}\frac{9N^5-10N^4-11N^3+68N^2+24N+16}
                            {(N-1)N^2(N+1)^2(N+2)^2}{\zeta_3}
                 -\frac{P_3}
                         {(N-1)N^3(N+1)^3(N+2)^3}{S_2}
\nonumber\\  &&                  
 -\frac{2P_{4}}
                          {(N-1)N^3(N+1)^3(N+2)^2}{\zeta_2}               -\frac{P_5}
                         {N(N+1)^3(N+2)^3}{S^2_1}
\nonumber\\ & &
+\frac{2P_{6}}
                          {N(N+1)^4(N+2)^4}{S_1}
                   -\frac{2P_{7}}
                          {(N-1)N^5(N+1)^5(N+2)^5}
                \Biggr\}~.\label{aQg2bar}
\nonumber
\end{eqnarray}
}
\includegraphics[]{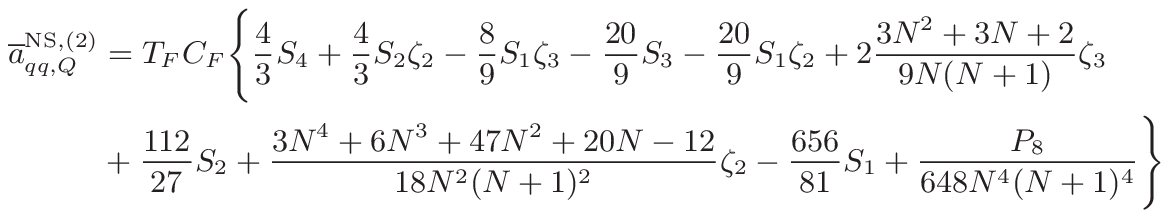}\\[0.5em]
\includegraphics[]{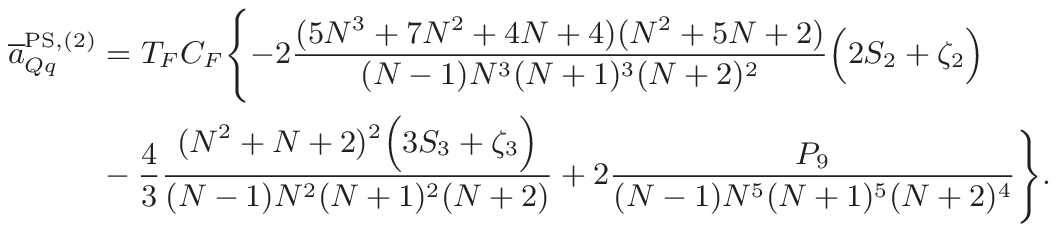}\\[0.5em]
with the polynomials $P_i$ given by\\[0.5em]
\includegraphics[]{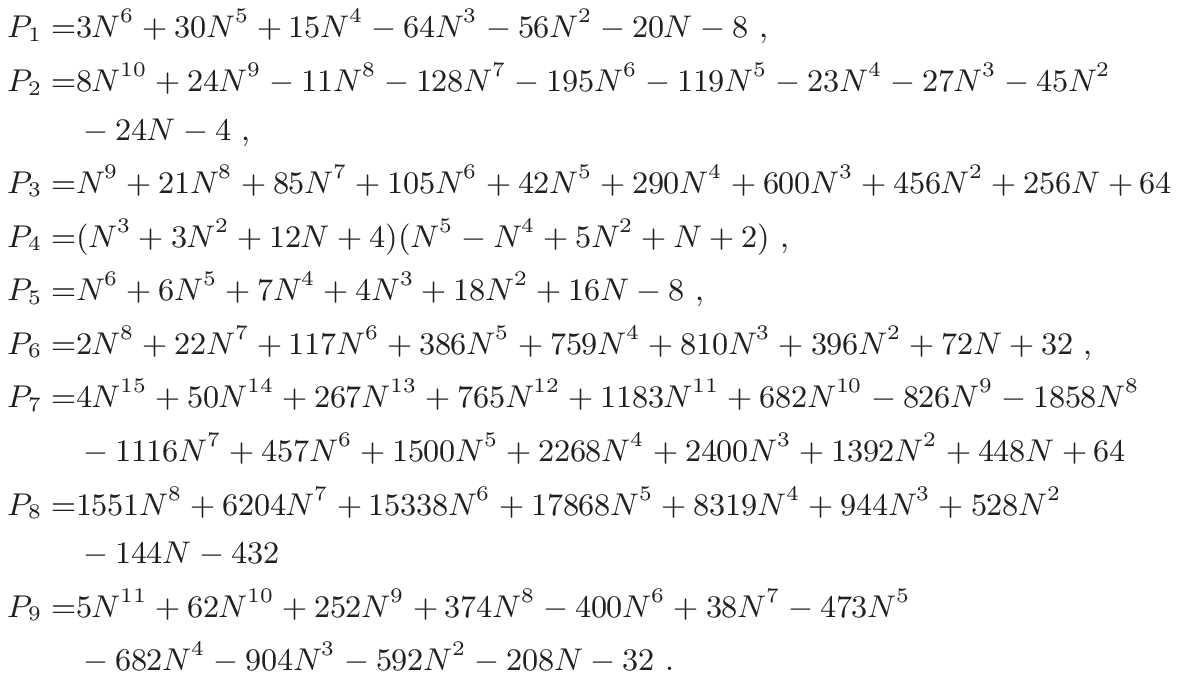}\\[1em]
Here, $S_{i,...,j}\equiv S_{i,...,j}(N)$, $\beta \equiv \beta(N+1)$,
and $\zeta_i \equiv \zeta(i)$ is Riemann's Zeta-function. In all
results, there is an overall factor
$S_{\eps}^2a_s^2(m^2/\mu^2)^{\eps}~$ and an overall factor
$(1+(-1)^N)/2~$ in all singlet and pure--singlet cases.
The above Mellin-space expressions can be converted to $x$-space using
analytic continuation in $N$ and Mellin inversion \cite{ANCONT}.

\section{Comparison}
\noindent
The results in \cite{Buza:1995ie}, which are up to constant order in
$\eps$, involved 48 basic functions, cf.  \cite{Bierenbaum:2007qe}.
Our results for the constant part of the OMEs gave raise to only six
harmonic sums, where five of them can be obtained from $S_1(N)$ through 
differentiation after analytic continuation and using algebraic 
relations. This leads to an amount of only two basic functions
\cite{Bierenbaum:2007pn}.
To order O($\eps$), we encounter the following 14 harmonic sums: $$
\{{S_1,~S_2,~S_3,~S_4,~S_{-2},S_{-3},S_{-4}}\},\quad S_{2,1},\quad
S_{-2,1},\quad S_{-3,1}, \quad S_{2,1,1},\quad S_{-2,1,1}, $$
$$
S_{-2,2}, \quad S_{3,1}. $$
We can again group the first seven
functions into the same class. Additionally, one finds that the
function ${S_{-2,2}}$ depends on ${S_{-2,1}}$ and ${S_{-3,1}}$, and
the harmonic sum ${S_{2,1}}$ depends on ${S_{3,1}}$, which leaves us
to order $\eps$ with only six basic harmonic sums, as also observed for a 
large variety of other two--loop processes, cf.~\cite{Blumlein:2007dj}. 
\section{Conclusion}
\noindent
We have calculated the O($\eps$)--term of the unpolarized massive
two--loop OMEs, contributing to the heavy--flavour Wilson coefficients
in the asymptotic limit $Q^2 \gg m^2$, as a first step towards the
O($\alpha_s^3$)--term of these Wilson coefficients. The calculation
was done in Mellin space, where numeric results were obtained by the
use of Mellin--Barnes integrals, whereas analytic results were
calculated using generalized hypergeometric functions. After applying
algebraic relations, the analytic result for the O($\eps$)--term is
expressible in only six basic harmonic sums.

\vspace{1mm}
\noindent
{\bf Acknowledgement.}\\
This work was supported in part by DFG Sonderforschungsbereich
Transregio 9, Computergest\"utzte Theoretische Physik. 
We would like to thank C.~Schneider for useful comments.


\end{document}